\documentclass[12pt]{article}
\usepackage{geometry}
\usepackage{scrextend}
\usepackage[english]{babel}
\usepackage[utf8]{inputenc}
\usepackage{amsmath, amssymb}
\usepackage{graphicx}
\usepackage{subcaption}
\usepackage{float}
\usepackage[usenames, dvipsnames]{xcolor}
\usepackage{mathrsfs}
\usepackage[colorinlistoftodos]{todonotes}

\newcommand{\margin}{\begin{addmargin}[1em]{2em}}
\newcommand{\emargin}{\end{addmargin}}

\usepackage{bm}
\usepackage{cite}
\usepackage{enumitem}
\usepackage[linktoc=page,
      colorlinks=true,
      linkcolor=blue,
      citecolor=blue,
      urlcolor=black,
      filecolor=blue,
      pdfstartview=FitV,
      pdfpagemode=None,
        bookmarksopen=true
      ]{hyperref}

\title{Marginally Trapped Surfaces and AdS/CFT}
\author{Brianna Grado-White and Donald Marolf\\\small{\textit{Department of Physics, University of California, Santa Barbara, CA 93106, USA}}\\\small{\textit{Email: brianna@physics.ucsb.edu, marolf@physics.ucsb.edu}}}

\begin{document}
\maketitle
\begin{abstract}
It has been proposed that the areas of marginally trapped or anti-trapped surfaces (also known as leaves of holographic screens) may encode some notion of entropy. To connect this to AdS/CFT, we study the case of marginally trapped surfaces anchored to an AdS boundary. We establish that such boundary-anchored leaves lie between the causal and extremal surfaces defined by the anchor and that they have area bounded below by that of the minimal extremal surface. This suggests that the area of any leaf represents a coarse-grained von Neumann entropy for the associated region of the dual CFT. We further demonstrate that the leading area-divergence of a boundary-anchored marginally trapped surface agrees with that for the associated extremal surface, though subleading divergences generally differ. Finally, we generalize an argument of Bousso and Engelhardt to show that holographic screens with all leaves anchored to the same boundary set have leaf-areas that increase monotonically along the screen, and we describe a construction through which this monotonicity can take the more standard form of requiring entropy to increase with boundary time. This construction is related to what one might call future causal holographic information, which in such cases also provides an upper bound on the area of the associated leaves. 
\end{abstract}

\tableofcontents 
\section{Introduction}
One of the central and most striking pillars of black hole thermodynamics is the Bekenstein-Hawking entropy formula, which relates the entropy of a black hole to the area of its event horizon \cite{Hawking:1974, Bekenstein:1974}. The notion that the microscopic degrees of freedom of a spacetime are controlled by degrees of freedom on a codimension-one surface is known as holography, and the area/entropy formula has been generalized in various directions. One such generalization comes from the AdS/CFT correspondence in the form of the Ryu-Takayanagi formula (or its covariant generalization by Hubeny, Rangamani, and Takayanagi [HRT])  \cite{Maldecena:1998, RyuTakayanagi:2006, HubenyRangamaniTakayanagi:2007}. This formula relates the entropy of the dual CFT in some boundary region with the area of an extremal surface through the bulk. A second generalization comes from the conjectured Covariant Entropy Bound (or Bousso bound) which states that the area of an achronal codimension-two surface $\sigma$ bounds some entropy flux through any orthogonal null congruence to $\sigma$ having non-positive expansion $(\theta \leq 0)$. As noted in \cite{Bousso:1999}, marginally trapped surfaces have a special status with respect to this bound, as the two future-directed congruences have $\theta=0$ and $\theta \leq0$. When the Null Curvature condition holds, the bound would then apply to both congruences (and both future and past directions of the $\theta=0$ congruence). Such surfaces can be stitched together to form a continuous codimension-one surface, called a holographic screen, in which case the marginally trapped surfaces are called leaves of the screen. It was further shown in \cite{EngelhardtBousso:2015,EngelhardtBousso:2015-2} that the leaves of the holographic screen obey a monotonic area law, and thus presumably a thermodynamic second law.  

Despite this thermodynamic property, the nature of the entropy described by holographic screens has remained unclear.\footnote{During the preparation of this work, \cite{EngelhardtWall:2017} appeared which further clarified this issue.  See Section \ref{sec:disc} for further comments on \cite{EngelhardtWall:2017}.} In contrast, the Ryu-Takayanagi formula computes the von Neumann entropy, $\text{tr} (\rho \ \text{log} \ \rho)$, of the dual CFT density matrix $\rho$. A natural question, then, is whether these two notions can be connected. A first step in this direction is to notice that the extremal surface used by RT has vanishing expansions in both of its orthogonal null directions. Though the usual definitions of marginally trapped surfaces require that they be compact, if we generalize to the non-compact case the extremal surface can be thought of as a leaf of a holographic screen.\footnote{The connection between AdS/CFT holography and the Covariant Entropy Bound was explored in \cite{HubenyRangamaniTakayanagi:2007}.} This suggests that it may be useful to define a general notion of non-compact holographic screen anchored to appropriate boundary sets $\partial A$ on an asymptotically locally anti-de Sitter (AlAdS) boundary. 

We explore the properties of such screens below when all leaves are anchored to the same boundary set $\partial A$, where as for RT/HRT we take $\partial A$ to be the boundary of a partial Cauchy surface $A$ for the boundary spacetime. In contrast, as can be seen by considering screens where every leaf is an extremal surface, letting the anchor set vary from leaf to leaf would generally result in infinite area-differences of either sign between between nearby leaves, so such screens do not appear to satisfy a useful second law of thermodynamics. However, many of our results would nevertheless apply to that case as well.

We begin in Section \ref{sec:prelim} with a brief review and discussion of the method we will use to construct marginally trapped surfaces anchored to the AdS boundary. Section \ref{sec:order} then shows that, with certain assumptions, a marginally trapped surface must lie between the extremal surface and causal surface anchored to the same boundary region; i.e., that it lies inside the entanglement wedge but outside the causal wedge. We further show that the area of the marginally trapped surface equals or exceeds that of the corresponding extremal surface, suggesting that it describes a coarse graining of the von Neumann entropy. In addition, when a marginally-trapped surface anchored at $\partial A$ lies in the past horizon defined by an appropriate boundary region $S$ (with $\partial S = \partial A$), a construction naturally called future causal holographic information also gives on upper bound on the marginally-trapped area. Section \ref{sec:div} studies divergences in the area of the marginally trapped surfaces associated with the region near the AdS boundary and shows that, while the leading order divergences of our marginally trapped surfaces match those of the extremal surface, the subleading divergences generally differ. Section \ref{sec:thermo} then generalizes the thermodynamic results of \cite{EngelhardtBousso:2015, EngelhardtBousso:2015-2} to holographic screens with non-compact leaves. We close with some brief discussion in Section \ref{sec:disc}. In particular, for surfaces on the past horizon of a boundary set $S$ as above, we note that the results of Section \ref{sec:thermo} can take the form of a standard second law in that they imply non-decrease in area under arbitrary deformations of $S$ toward the future, so long as $\partial S$ remains fixed and the holographic screen moves in a spacelike direction.

\section{Preliminaries}
\label{sec:prelim}

This section provides some definitions and lemmas that will be used throughout the work below. It also summarizes the method we use to construct boundary-anchored holographic screens and thus defines the class of such surfaces to be studied.  

We assume that the bulk spacetime obeys the Null Curvature Condition, $R_{ab}k^ak^b \geq 0$ for any null vector $k^a$ and is AdS globally hyperbolic. The latter condition (see e.g. \cite{EngelhardtWall:2013}) means that there is an achronal surface $\Sigma$ for which the AdS-domain of dependence $D(\Sigma) = D^+(\Sigma) \cup D^-(\Sigma)$ is the entire spacetime. Here $D^+(\Sigma)$ ($D^-(\Sigma)$) is the set of points $p$ for which all past-inextendible (future-inextendible) causal curve through $p$ intersects either $\Sigma$ or the AlAdS boundary.  \\

\textit{Definition:} A \textit{future holographic screen} $H$ is a smooth hypersurface which admits a foliation by marginally trapped surfaces, called \textit{leaves}. A \textit{marginally trapped surface} is a smooth, codimension-two achronal spacelike surface whose future directed orthogonal null congruences, $k$ and $\ell$, have expansions satisfying
\begin{equation}
\begin{aligned}
\theta_k &= 0,\\
\theta_{\ell} &\leq 0. 
\end{aligned}
\end{equation}
Similarly, we can define a \textit{past holographic screen} as a smooth hypersurface which admits a foliation by marginally \textit{anti-trapped surfaces}, so that $\theta_l\geq 0$. Note that an extremal surface will have $\theta_{\ell} = \theta_k=0$. \\

Holographic screens are also known as marginally trapped tubes \cite{AshtekarKrishnan:2003}, and are a generalization of dynamical horizons, removing the restriction \cite{EngelhardtBousso:2015} that the surface be spacelike. Since we focus on the boundary-anchored case, we also omit the usual requirement that the marginally trapped surfaces be compact. In addition, we require such boundary-anchored marginally trapped surfaces $\sigma$ to be homologous to some partial Cauchy surface $A$ for the boundary spacetime. Here by `homologous to $A$', we mean that there is a bulk AdS-Cauchy surface $\Sigma = \Sigma_1 \cup \Sigma_2$ with $\partial \Sigma = \sigma \cup A$. As a result, $\partial \sigma = \partial A$. In this work we use the symbol $\partial X$ to denote the boundary of any set $X$ as computed in the conformal compactification of our AlAdS spacetime; i.e., $\partial X$ will include any limit points of $X$ in the AlAdS boundary. In contrast, we will use the notation $\dot{X}$ to refer to the boundary of $X$ as defined by the natural topology of the bulk spacetime in which the bulk is an open set. As a result, $\dot{X}$ cannot intersect the AlAdS boundary, but $\partial \dot{X} :=  \partial(\dot{X})$ contains precisely those points in the AlAdS boundary which are limit points of $\dot{X}$.

We will focus in particular on future holographic screens where, for the boundary-anchored case, we define the $k$, $\ell$ null congruences as follows: consider a boundary region $A$ and a marginally trapped surface $\sigma$ homologous to $A$ as above. We define the $k$ null congruence orthogonal to $\sigma$ to be the one launched towards the future from the $\Sigma_1$ side of $\sigma$, while the $l$ null congruence orthogonal to $\sigma$ is the one launched toward the future from the $\Sigma_2$ side. Note that AdS-global hyperbolicity requires $\dot{D}^+(\Sigma_1)\backslash\Sigma_1 = \dot{I}^+(\Sigma_2)$, and in fact $\dot{D}^+(\Sigma_1)\setminus \Sigma_1 \subset \dot{I}^+(\sigma)$, which implies that it is generated by the $k$-congruence just defined.  

\begin{figure}[h]
\centering
\includegraphics[width = 0.5\linewidth]{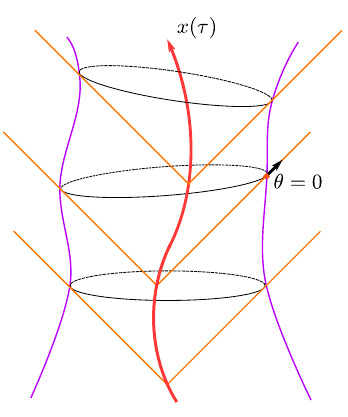}
\caption{A holographic screen can be constructed by null foliation from an observer's light cones. For example, given an spacetime path $x(\tau)$, on each future lightcone emanating from that path, we can find the maximal area codimension-two surface, if it exists. This surface will then have $\theta = 0$ along the lightcone. Stitching together all these surfaces on all of the future lightcones associated with the observer will give us our holographic screen.}
\label{fig:nullfol}
\end{figure}

A well known property of holographic screens is that they are highly non-unique: changing the foliation of the spacetime generally changes the holographic screen (see, e.g. \cite{EngelhardtBousso:2015-2}). Previous work has focused on generating them from null foliations (e.g. \cite{EngelhardtBousso:2015,EngelhardtBousso:2015-2, SanchesWeinberg:2016}), building the leaves of the holographic screen by finding a codimension-two surface with maximal area on each null slice (see Figure \ref{fig:nullfol}). In the case where the null foliations are taken to be the set of past or future light cones emanating from an observer's worldine, the foliation dependence of screens can be thought of as an observer dependence.

However, if we were to use this null construction in empty AdS, then the maximal cut of any null surface would lie on the AdS boundary. The holographic screen would then just be the usual conformal boundary of the spacetime. This is consistent with the Bousso-bound picture, in that the degrees of freedom in the boundary CFT control the bulk degrees of freedom, but seems rather trivial. In particular, even the renormalized area is strictly infinite. 

\begin{figure}[h]
\centering
\includegraphics[width=.37\linewidth]{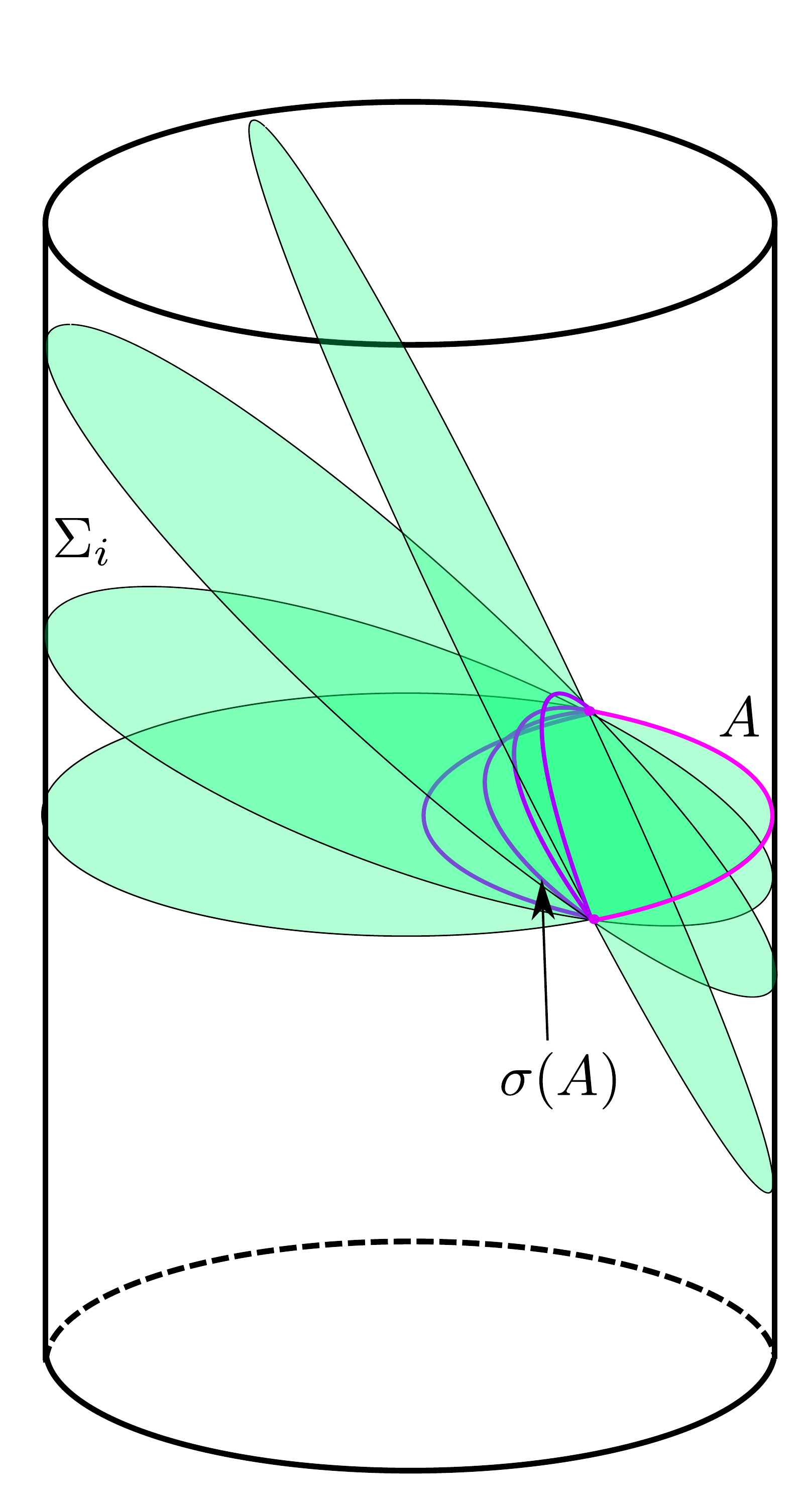}\quad%
\caption{For a fixed boundary region $A$, we construct our holographic screens by first fixing a foliation (pictured here as green slices), such that each slice contains $\partial A$ (where $A$ is the pink section). Then, on each slice, we solve for the trapped surface (the purple curves). Each trapped surface is a leaf of a holographic screen, and stitched together, the collection of leaves comprises our holographic screen. In practice, we find that solutions typically exist. The figure is based on numerical results in Schwarzschild-AdS.}
\label{construction}
\end{figure}

For a fixed subset $A$ of some boundary Cauchy surface $C$, we instead wish to construct a marginally trapped surface through the bulk and anchored to the boundary $\partial A$ of that region. To do so, instead of using a null foliation as above, we pick any foliation of our bulk spacetime such that each slice $\Sigma_i$ contains $\partial A$. On $\Sigma_i$, we can then attempt to solve for a marginally trapped surface also anchored to $\partial A$, giving us our leaf $\sigma_i(A)$. See Figure \ref{construction} for a depiction. Although we leave a complete analysis for future investigation, in practice we find that solutions exist. While the leaves of the screen are required to be spacelike, the same need not be true of the slices $\Sigma_i$ used to construct them.

Indeed, we solve for our leaves in the following manner (see \cite{HubenyRangamaniTakayanagi:2007} for a similar setup): a general codimension-two surface $S$ can be specified by two constraints, 
\begin{equation}
\begin{aligned}
F(x^\mu)&=0,\\
G(x^\mu)&=0.
\end{aligned}
\end{equation}
The gradients $\nabla_\nu F(x^\mu)$ and $\nabla_\nu G(x^\mu)$ are then vectors orthogonal to $S$. When they are independent and $S$ is spacelike, we can write the orthogonal null vectors as some linear combinations
\begin{equation}
n_{\nu,a}= \nabla_\nu F(x^\mu) + c_a\nabla_\nu G(x^\mu),
\end{equation}
for $c_a$ constants and $a=\ell,k$ with $k_\mu = n_{\mu,k}, \ell_\mu = n_{\mu,\ell}$. The null extrinsic curvatures are then
\begin{equation}
\chi_{\mu \nu ,a} = {\tilde{g}^\rho}_\mu {\tilde{g}^\lambda}_\nu \nabla_\rho n_{\lambda, a},
\end{equation}
where $\tilde{g}$ is the induced metric on $S$:
\begin{equation}
\tilde{g}_{\mu \nu} = g_{\mu\nu} + \ell_{\mu} k_{\nu} +  \ell_{\mu} k_{\nu}.
\end{equation}
Finally, each expansion is the trace of the appropriate null extrinsic curvature:
\begin{equation}
\theta_a = {\chi^\mu}_{\mu ,a}.
\end{equation}

The extremal surface anchored to $\partial A$ is then found by solving $\theta_k = \theta_\ell =0$ to find $F$ and $G$. But for our marginally trapped surfaces, only $\theta_k$ need vanish so the solution is underdetermined. We may hope to specify a unique solution by taking $G(x^\mu) = t - \hat{G}(x^i)=0$, for some particular $\hat{G}$ (with $\{x^\mu\} = \{t, x^i\}$), and to then solve $\theta_k=0$ for $F$.

Once we have found our holographic screen, we will want to compare it to both the causal wedge and entanglement wedge as defined below (following \cite{HubenyRangamaniTakayanagi:2007}).\\

\textit{Definition:}  For a given boundary region $A$, we will denote the boundary domain of dependence by $D_{bndy}(A)$. The \textit{causal wedge} is then defined as the intersection of the bulk past and future of this domain of dependence, $\mathcal{C}(A) = I^-(D_{bndy}(A)) \cap  I^+(D_{bndy}(A))$. The \textit{causal information surface} or \textit{causal surface} $\Xi_A$ lies on the boundary of this region, and is given by the intersection of the past and future bulk horizons of the boundary domain of dependence of $A$, $\Xi_A = \dot{I}^-(D_{bndy}(A)) \cap  \dot{I}^+(D_{bndy}(A))$. \\

\textit{Definition:}  Let the HRT surface $m(A)$ be the codimension-two surface with extremal area in the bulk, anchored to the boundary $\partial A$ of $A$. We also require that $m(A)$ be homologous to $A$ in the sense discussed above for marginally trapped surfaces. If there are multiple extremal surfaces satisfying this constraint, take the one with least area. The entanglement wedge $\mathcal{E}(A)$ is then the bulk AdS-domain of dependence $D(\Sigma)$ of any partial AdS-Cauchy surface $\Sigma$ satisfying $\partial \Sigma = A \cup m(A)$.  \\

We can also define a similar wedge $\mathcal M(\sigma)$ associated with any marginally trapped surface.\\ 

\textit{Definition:} For any marginally trapped surface $\sigma$ homologous to $A$ we define the \textit{marginally trapped wedge} ${\cal M}(\sigma)$ to be the bulk domain of dependence $D(\Sigma)$ of any partial AdS-Cauchy surface $\Sigma$ satisfying $\partial \Sigma = \sigma \cup A$.  \\ 

\begin{figure}
\centering
\begin{subfigure}{.47\linewidth}
\centering
\includegraphics[width=0.96\linewidth]{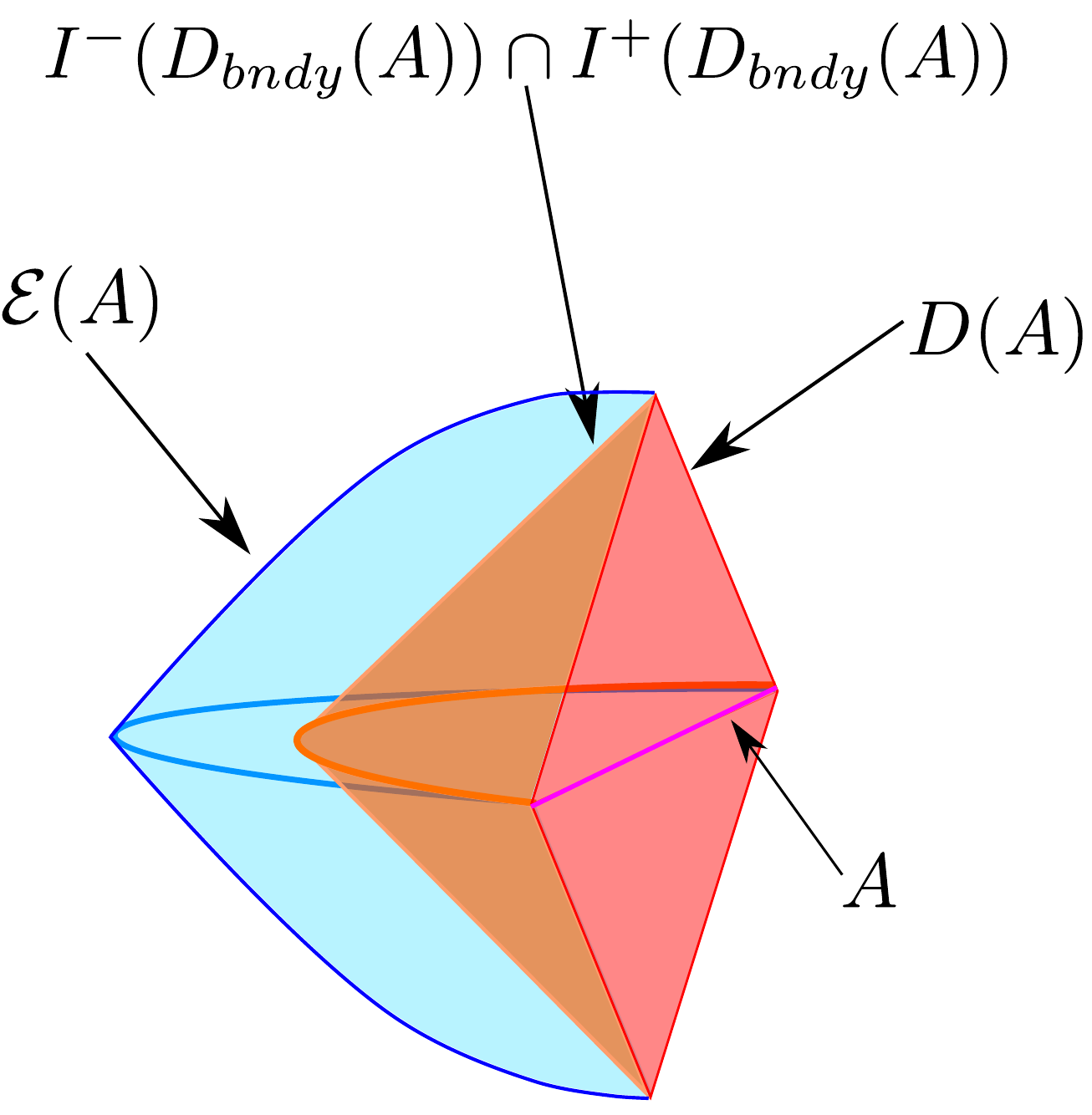}\quad%

\end{subfigure}
\begin{subfigure}{.47\linewidth}
\centering
\includegraphics[width=0.96\linewidth]{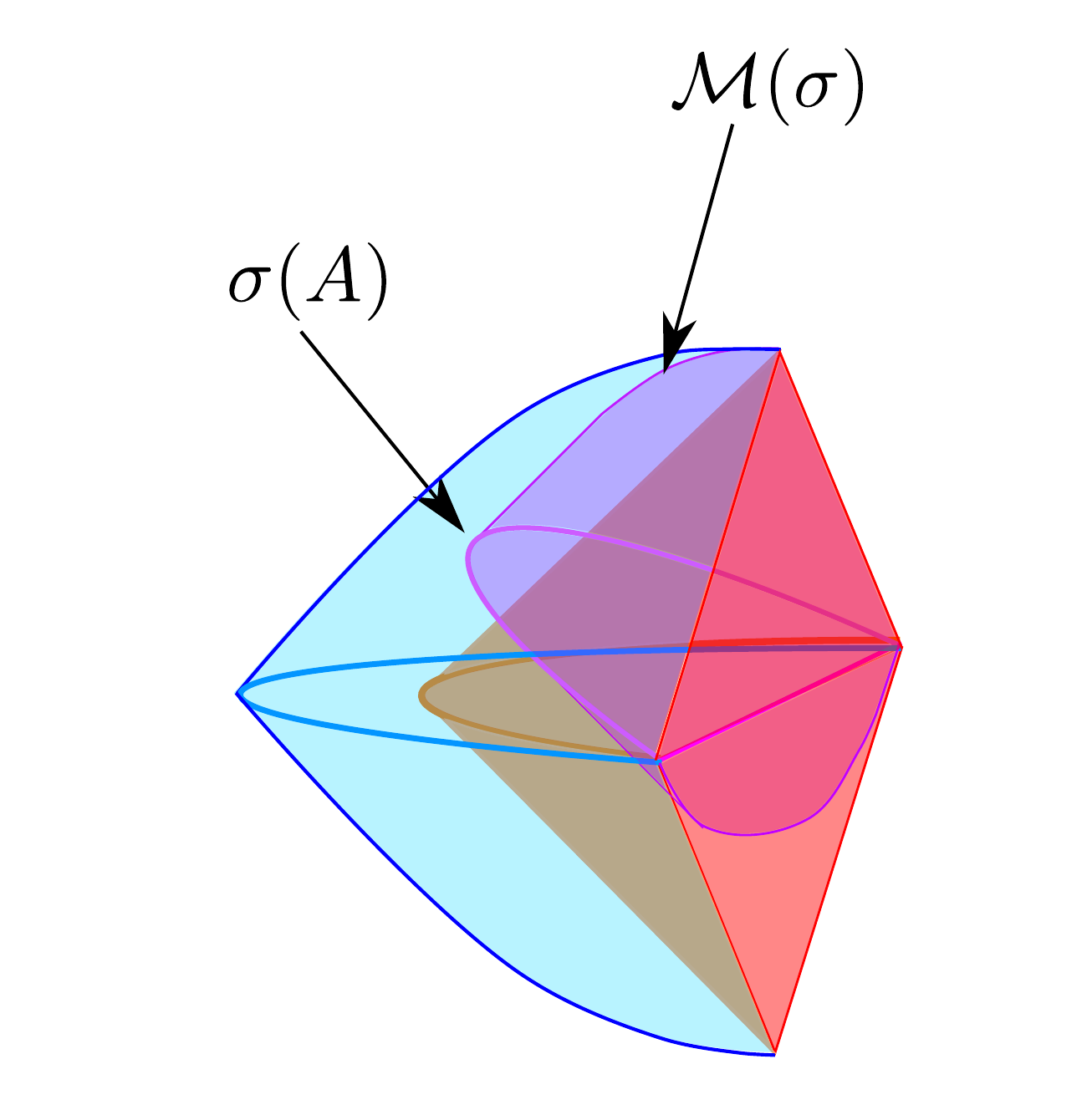}\quad%

\end{subfigure}
\caption{{\bf Left:} Depiction of the Entanglement Wedge (blue) and Causal Wedge (orange). The entanglement wedge $\mathcal{E}(A)$ is the domain of dependence lying between the extremal surface $m(A)$ and the boundary. In contrast, the causal wedge $I^-(D_{bndy}(A))\cap I^+(D_{bndy}(A))$ is defined as the intersection of the bulk past and future of the domain of dependence $D_{bndy}(A)$ in the boundary. The intersection of the past and future horizons defines the causal surface. {\bf Right:} Depiction of the marginally trapped surface $\sigma(A)$ which (as shown in Section \ref{sec:order}) must lie in the entanglement wedge but above the future horizon of $D_{bndy}(A)$. The associated marginally trapped wedge $\mathcal{M}(\sigma)$ is also shown (purple).}
\end{figure}

In addition to the above definitions, we will repeatedly use the following Lemma.\\  

\textit{Lemma 2.1:} (From \cite{Wall:2012}) Suppose $N_1$ and $N_2$ are two null hypersurfaces that are tangent at some point $x$ on some slice $\Sigma$. Then if there exists some neighborhood of $x$ on $\Sigma$,  such that $N_2$ is nowhere to the past of $N_1$, then $\theta_{N_2}\geq \theta_{N_1}$ at $p$. \\

\begin{figure}[h]\label{fig:nullcongruence}
\centering
\includegraphics[width=.4\linewidth]{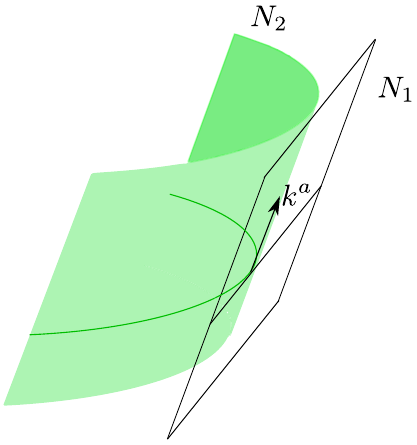}\quad%
\caption{The lines depict intersecting spacelike cuts of null congruences $N_1$ and $N_2$. $N_2$ is nowhere to the past of $N_1$ and is thus expanding faster by Lemma 2.1.}
\end{figure}

This Lemma is especially useful when combined with the following result (often left implicit in applications of Lemma 2.1).\\

\textit{Lemma 2.2} If a smooth spacelike curve $\gamma$ intersects the boundary of the future $I^+(S)$ of some set $S$ at a point $p$, then either i) $\gamma$ enters the chronological future $I^+(S)$ or ii) all null generators of $\dot{I}^+(S)$ through $p$ intersect $\gamma$ orthogonally. 

\textit{Proof:} By e.g. Theorem 8.1.6 of \cite{Wald:1984}, $p$ lies on a null geodesic $\lambda$ (perhaps with a past endpoint) that (at least to the past of $p$) lies entirely in $\dot{I}^+(S)$. Let $k^a$ and $\zeta^a$ be vectors respectively tangent to $\lambda$ and $\gamma$ at $p$. Since $k^a$ is null and $\zeta^a$ is spacelike, then either a) $k^a$ and $\zeta^a$ span a timelike plane or b) $k^a$ and $\zeta^a$ span a null plane, and are orthogonal. So if any null generator $\lambda$ through $p$ fails to intersect $\gamma$ orthogonally, case (a) must hold for that generator. We can then find a local Lorentz frame where $k^a\partial_a \propto \partial_t + \partial_x$ and $\zeta^a \propto \partial_x$, so $\gamma$ clearly enters $I^+(S)$.\\

Combining Lemmas 2.1 and 2.2 gives us the following: \\

\textit{Corollary 2.3} If a codimension-two surface $\sigma$ intersects $\dot{I}^+(S)$ at $p$ then either $\sigma$ enters $I^+(S)$ or every generator $\lambda$ of $I^+(S)$ at $p$ has a well-defined expansion $\theta^{\lambda}(p)$
 that is equal to or greater than the expansion $\theta^{\sigma}(p)$ along the same null geodesic as defined by the associated null congruence orthogonal to $\sigma$.
 
\textit{Proof:} If $\sigma$ fails to enter $I^+(S)$, then by Lemma 2.2, all null generators of $I^+(S)$ at $p$ intersect $\sigma$ orthogonally. Consider such a generator $\lambda$, together with the nearby generators in $\dot{I}^+(S)$. Unless $\lambda$ has a conjugate point at $p$, at least in a neighborhood of $p$ these generators define a smooth null surface $\mathcal{N}$ nowhere to the past of $\sigma$. As a result, in a neighborhood of $p$ the orthogonal null congruence to $\sigma$ containing $\lambda$ is also smooth and lies nowhere to its future. Thus $\theta^{\mathcal{N}}(p)\geq \theta^{\sigma}(p)$ by Lemma 2.1. Furthermore, if the expansion along $\lambda$ at $p$ is ill-defined (i.e. if $p$ is a conjugate point of $\lambda$), then any point $q \in \lambda$ to the future of $p$ also lies in the chronological future of geodesic generators $\lambda'$ of $I^+(S)$ close to $\lambda$, and which differ from $\lambda$ only to first order in $q-p$. As a result, such geodesics $\lambda'$ also lie in front (i.e. to the past) of the infinitesimal null plane $\mathcal{N}$ defined by $\lambda$ by an amount that is first order\footnote{We believe this to be true. What is straightforward to argue rigorously is that if the expansion of $\dot{I}^+(S)$ remains bounded when approaching $p$ from any direction, then we may argue just as in the case where $p$ is not a conjugate point. If a direction-dependent limit diverges along some spacelike cut, then the cut must deviate from $\mathcal{N}$ strictly faster than quadratically. Furthermore, it must do so by lying in front (to the past) of $\mathcal{N}$ so that $\lambda$ can enter its future immediately after the conjugate point at $p$. As a result, since $\sigma$ is tangent to $\mathcal{N}$, it also lies to the past of $\sigma$. I.e., $\sigma$ enters $I^+(S)$.} in the separation $\delta \lambda$ between $\lambda$ and $\lambda'$. But since $\sigma$ is smooth, it can bend in front of this null plane only at second order. Thus $\sigma$ enters the future of some $\lambda'$ and thus enters $I^+(S)$.

\section{Ordering of Surfaces}
\label{sec:order}

The RT and HRT surfaces measure the fine-grained entropy of the dual CFT in the associated domain. In addition, the Causal Surface has been conjectured to give a coarse-grained measure of the entropy known as the Causal Holographic Information \cite{HubenyRangamani:2012}; see related discussion in \cite{FreivogelMosk:2013,Kelly:2013aja,Engelhardt:2017wgc}. The idea that the latter is a coarse-graining of the former is associated with the fact that the causal surface lies closer to the boundary and has larger area than the extremal surface anchored to the same region \cite{HubenyRangamaniTakayanagi:2007,Wall:2012}.

In this section, we argue that any marginally trapped surface $\sigma$ anchored to the boundary at $\partial A$ lies in some sense between the above two surfaces. Specifically, we show it to lie inside the entanglement wedge but above the future horizon associated with $A$. We also show the area of $\sigma$ to be bounded below by the area of the extremal surface and -- in certain cases-- bounded above by the area of a cut of the causal horizon associated with future causal holographic information (fCHI).  So in such cases we expect the area of $\sigma$ to describe a coarse-grained entropy for the dual CFT that is finer than the coarse-graining associated with fCHI.

The proofs regarding the ordering of wedges are similar to proofs in \cite{Wald:1984, Wall:2012}. For arguments in Section 3.2, we assume that our marginal surface $\sigma$ can be approximated by a sequence $\sigma_i$  of surfaces anchored to the same $\partial A$, lying in a common AdS-Cauchy surface $\Sigma$ in which all of the surfaces $\sigma, \sigma_i$ are homologous to $A$, and maintaining $\theta_{\ell}\leq 0$ but having $\theta_k>0$. We can call such $\sigma$ \textit{deformable}, indicating that they may be deformed to cases with $\theta_k>0$, $\theta_{\ell}\leq 0$.

\subsection{Leaves Lie Outside of the Causal Wedge}
Our first result generalizes the well-known theorem that apparent horizons lie to the future of event horizons \cite{Wald:1984}. 

\textit{Theorem 3.1.1:} Let $\sigma$ be a marginally trapped surface, anchored to a boundary region $A$. Then it will lie above the future horizon of $D_{bndy}(A)$, and in particular outside the causal wedge $\mathcal{C}(A)$.

\textit{Proof:} We assume $A$ to be connected, as otherwise we can simply work with each connected component. If $\sigma$ fails to lie above the future horizon of $D_{bndy}(A)$, then some $p \in D_{bndy}(A)$ lies in the future of $\sigma$. But by the homology constraint, $\sigma$ lies on a Cauchy surface $\Sigma$ containing $A$, so $A$ is not in the future of $\Sigma$. Thus, there are points in $D_{bndy}(A)$ that are not to the future of $\sigma$. But since $A$ is connected, so is $D_{bndy}(A)$, and so some $q \in D_{bndy}(A)$ must lie on the boundary of the future of $\sigma$. Since $q$ is in the interior of $D_{bndy}(A)$, there is an open set $U \ni q$, $U \subset D_{bndy}(A)$ that does not intersect the future of $\partial A$. As a result, the closure $K$ of the set of points $r \in \sigma$ that can send future-directed timelike curves to $U$ is compact. 

Thus $q$ lies on the boundary of the future of $K$. Since $K$ is compact, this means there is a null generator $\lambda$ of $\dot{I}^+(K)$ that reaches $q$, and which in particular reaches the AlAdS boundary. Thus $\lambda$ maintains $\theta=0$ for infinite affine parameter. It follows that adding any perturbation which makes all null generators of the $k$-congruence from $\sigma$ satisfy the generic condition (i.e. that there exists non-vanishing null-curvature or shear along any segment of any null congruence) moves $\partial I^+(\sigma)$ off of $D_{bndy}(A)$. For example, we can throw null particles into the bulk from $D_{bndy}(A)$ just below every point of $\partial I^+(\sigma)$. Then, however, such particles clearly intersect the generators of $\dot{I}^+(\sigma)$ near the AlAdS boundary and can only move $\partial I^+(\sigma)$ by a small amount. But this contradicts the fact that $D_{bndy}(A)$ is an open set, so a small change in $\partial I^+(\sigma)$ cannot in fact remove the intersection with $D_{bndy}(A)$. We thus conclude $I^+(\sigma) \cap D_{bndy}(A) = \emptyset$ so that no part of $\sigma$ is below the future horizon. \\

For a certain class of extremal surfaces, we can also use a cut of the causal horizon to bound the area of the marginally trapped surface in the following sense:\footnote{We thank Aron Wall for a discussion regarding this point.}

\textit{Theorem 3.1.2:} Suppose a marginally trapped surface $\sigma$ anchored to $\partial A$ lies on the boundary of $I^+(S)$ of some $S$ in the AlAdS boundary for which $\partial S = \partial A$. Then the boundary of $I^+(S)$ will also intersect the future causal horizon defined by $D_{bndy}(A)$ in some cut $Y$, and the generators of $\dot{I}^+(S)$ define an area non-decreasing map from $\sigma$ into $Y$.

\textit{Proof:} Note that since the expansion of $\dot{I}^+(S)$ vanishes on the AlAdS boundary it is negative or zero everywhere on $\dot{I}^+(S)$. By Theorem 3.1.1, any generator of $\dot{I}^+(S)$ that reaches $\sigma$ does so after (or simultaneously with) passing through $Y$. Thus the map defined by these generators from $\sigma$ to $Y$ cannot decrease local areas. \\

The surface $Y$ that gives the bound was introduced in \cite{Kelly:2013aja} as a modification of causal holographic information conjectured to be associated with the future boundary domain of dependence $D^+_{bndy}(Y)$. The quantity $A/4G$ for $Y$ is thus naturally called future causal holographic information.

Now, Theorem 3.1.2 provides a sense in which the area of $\sigma$ is bounded below by that of $Y$. However, both areas are infinite, and we are typically interested in either the renormalized area (defined by subtracting an appropriate set of counter-terms) or the regularized area (defined by imposing a cut-off $z=z_0$ using some Fefferman-Graham coordinate $z$) in the limit $z_0 \rightarrow 0$ in which the regulator is removed. Using Theorem 3.1.2 to compare renormalized or regulated areas thus requires understanding how many generators of $\dot{I}^+(S)$ cross the regulator surface $z=z_0$ between $Y$ and $\sigma$ in the limit $z_0 \rightarrow 0$. The limiting flux of such generators is known to be finite \cite{Bunting:2015sfa} when the boundary of $D_{bndy}(S)$ is a boundary Killing horizon, but extrapolating those results to the more general case suggests that the flux generally diverges as $z_0^{-(d-2)}$ and that this divergence can take either sign. Indeed, the total area of $\dot{I}^+(S)$  lost through the $z=z_0$ regulator surface takes the form

\begin{equation}
\label{eq:LostArea}
{\rm Lost  \ Area} \sim \int_{Y}^{\sigma} d \lambda \int_{\partial A_{z_0}} l^{d-2} z^{-(d-2)} \sqrt{\tilde{q}^{0}} \frac{1}{z} \frac{\partial z}{\partial \lambda},  
\end{equation}
where $l$ is the AdS scale, $\partial A_{z_0}$ is a regulated version of $\partial A$ located at $z=z_0$, $\sqrt{\tilde{q}^{(0)
}}$ is the area element on $\partial A$ of the finite-but-unphysical metric on the AlAdS conformal boundary (see Section \ref{sec:leading}),  and $\lambda$ is a smooth parameter along each geodesics between $Y$ and $\sigma$. If $Y$ and $\sigma$ admit power series expansions in $z$  (perhaps with possible log terms at order $z^d$) we generally have 
$\frac{1}{z} \frac{\partial z}{\partial \lambda} \sim 1$, and also $\int_{Y}^{\sigma} d \lambda \sim z$ since $Y$ and $\sigma$ both intersect the AlAdS boundary at $\partial A$. The lost area is then $O(z^{-(d-3)})$, so we learn that the regulated/renormalized area of $Y$ exceeds that of $\sigma$ if these areas differ by a term more divergent than $z^{-(d-3)}$. In particular, the bound applies to the coefficient of the leading divergence at $O(z^{-(d-2)}$). But we learn nothing if the areas are already known to coincide to higher order.

On the other hand, we expect the case of most interest to occur when both $Y$ and $\sigma$ coincide asymptotically with the extremal surface $m(A)$ anchored at $\partial A$. Consider then the renormalized areas of $Y$ and $\sigma$ defined by subtracting the known counter-terms for extremal surfaces areas. Since the regulated areas (before subtracting counter-terms) are of the form

\begin{equation}
{\rm Regulated  \ Areas} \sim \int_{z_0} \frac{dz}{z} \int_{\partial A_{z_0}} l^{d-2}z^{-(d-2)} \sqrt{\tilde{q}^{(0)}},  
\end{equation}
the renormalized areas of $Y$ and $\sigma$ are generally finite only when these surfaces coincide with $m(A)$ up to corrections vanishing faster than $z^{d-2}$ by some power law. Comparing with \eqref{eq:LostArea} immediately yields the following result:

\textit{Theorem 3.1.3:} Suppose $\sigma$ and $Y$ in theorem 3.1.2 both agree with $m(A)$ up to corrections vanishing faster than $z^{d-2}$ by some power law.  Then the renormalized areas of $\sigma$ and $Y$ are finite, and the renormalized area of $Y$ equals or exceeds that of $\sigma$.


\subsection{Leaves Lie Inside of the Extremal Wedge}
We now show that boundary anchored marginally trapped surfaces lie inside the extremal wedge $\mathcal{E}(A)$, as long as there is an appropriate region through which we can deform extremal surfaces while keeping them extremal. This condition is related to the absence of extremal surface barriers as defined in \cite{EngelhardtWall:2013}. 

\textit{Theorem 3.2.1} Let $\sigma$ be a deformable marginally trapped surface anchored to $\partial A$, with a sequence of approximating surfaces $\sigma_i$. All $\sigma_i$ lie in some AdS-Cauchy surface $\Sigma$, such that $\partial \Sigma \supset A$. Suppose there is a one parameter family $m(I)$ of extremal surfaces\footnote{Note that a general $m(I)$ need not be an HRT surface, as it need not be the extremal surface of minimal area.} anchored to the boundary on non-overlapping sets $\partial A_I \subset \partial \Sigma$ such that i) m(I) is continuous in $I$ for $I \in \left[0,1\right]$, ii) $m(I=0) = m(A)$ but $\partial A_I \subset \partial \Sigma \setminus \overline{A}$. for $I>0$, iii) each $\sigma_i$ is contained in the extremal wedge associated with $m(I=1)$, and iv) each $\partial A_I$ is the boundary of some boundary set $A_I$ homologous to $m(I)$. Then $\sigma$ lies in the closure of the entanglement wedge $\mathcal{E}(A)$. 

\begin{figure}[h]\label{fig:extremalcongruence}
\centering
\includegraphics[width=.7\linewidth]{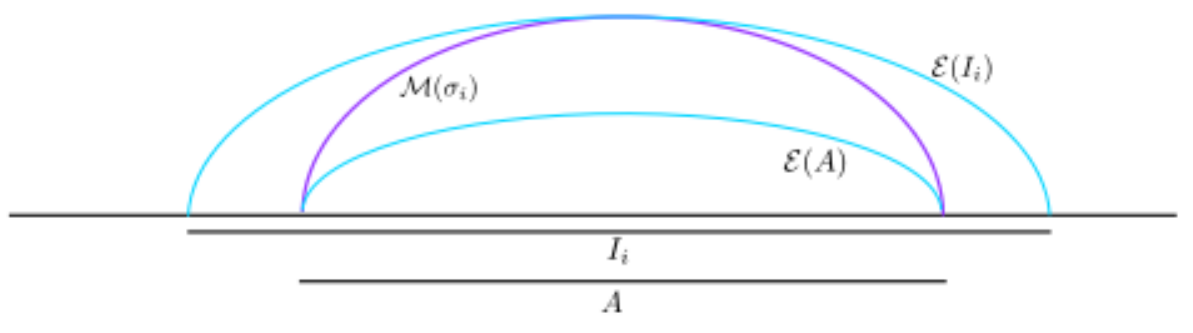}\quad%
\caption{If a marginally trapped surface (or one of its approximating curves) $\sigma_i(A)$ lies outside the corresponding entanglement wedge $\mathcal{E}(A)$, we can continuously deform our boundary to larger regions, until at some point the entanglement wedge just touches the wedge associated with $\sigma_i$. This results in a contradiction.}
\label{fig:def}
\end{figure}

\textit{Proof:} Define $\Sigma_{1i}$, $\Sigma_{2i}$ to be the regions in $\Sigma$ such that $\Sigma_{1i} \cup \Sigma_{2i} = \Sigma$ and $\partial \Sigma_{1i} = \sigma_i \cup A$. For $m(I)$, we similarly define $\Sigma_I$, $\Sigma_{1I}$, and $\Sigma_{2I}$. Note that we may choose $\Sigma_{1I}$ to be continuous in $I$.\footnote{This continuity is automatic unless there is a connected component of the bulk spacetime that does not have an AdS boundary; i.e. the bulk contains a closed cosmology in addition to the asymptotic AdS piece. Such cases do not appear to be allowed in AdS/CFT, but for completeness we include them here.} We can also define the wedges associated to each of the surfaces of interest: $\mathcal{E}(I) = D(\Sigma_{1I})$, and $\mathcal{M}(\sigma_i) = D(\Sigma_{1i})$. Let $I_i$ be the smallest $I$ for which the closure $\overline{\mathcal{E}(I)}$ of ${\mathcal{E}(I)}$ contains $\mathcal{M}(\sigma_i)$; see figure \ref{fig:def}. Then, since $\mathcal{E}(I)$ is continuous in $\Sigma_{1I}$ (and thus in $I$), if $I_i \neq 0$ there must be some point $p$ that lies in the boundaries of both $\mathcal{E}(I_i)$ and $\mathcal{M}(\sigma_i)$.  Note that in this case the point $p$ cannot lie on the AdS boundary since $\partial A_{I_i} \subset \partial \Sigma \setminus \bar A$.

Now, since $p \in \dot{\mathcal{M}}(\sigma_i)$, it is connected to $\sigma_i$ by a null geodesic $\lambda \subset \overline{\mathcal{M}(\sigma_i)}\subset \overline{\mathcal{E}(I_i)}$. But no point of $\lambda$ can lie in the interior of $\mathcal{E}(I_i)$, as then $p$ would also lie in the interior of $\mathcal{E}(I_i)$ and not on the boundary of $\mathcal{E}(I_i)$. So if $q$ is the past endpoint of $\lambda$ on $\sigma_i$ we must also have $q$ lying in the boundary of $\mathcal{E}(I_i)$.  Furthermore, it is clear that the $k$-congruence from $\sigma_i$ is the one that locally does not enter $I^+(\Sigma_{2I_i})$. By the null convergence condition and Corollary 2.3 we then have that the expansions through $q$ defined by the orthogonal null congruences from $m(I_i)$ and $\sigma_i$ satisfy $0 \geq \theta^{m(I_i)} \geq \theta^{\sigma_i} >0$. This is a contradiction, so  $I_i=0$ for all $i$.  In particular, $\sigma_i \subset \mathcal{E}(A)$ and thus $\sigma \subset \overline {\mathcal{E}(A)}$ as desired.

Note that condition (iii) that each $\sigma_i$ be contained in the wedge associated with $m(I=1)$ is realized whenever we can deform the boundary region $\partial A$ to a point through $\Sigma \setminus \bar A$.

We can also use the extremal surface anchored at $\partial A$ to bound the area of appropriate similarly-anchored marginally trapped surfaces. The useful notions of `appropriate' are defined by issues involving the regulator surfaces $z=z_0$ as in the discussion of Theorems 3.1.2 and 3.1.3.  

\textit{Theorem 3.2.2:} Given a marginally-trapped surface $\sigma$ and an HRT surface $m(A)$ both anchored to $\partial A$, the renormalized area $\text{Area}_{\text {ren}}$ of $\sigma$ equals or exceeds that of $m(A)$ if i) $\partial A$ lies on a Killing horizon of the boundary or ii) $\sigma$ coincides with $m(A)$ up to corrections vanishing faster than $z^{d-2}$ by some power law. More generally, the coefficient of the leading divergence in the area of $\sigma$ equals or exceeds that for $m(A)$.

\textit{Proof:} We can use the maximin construction of HRT surfaces \cite{Wall:2012} to find an AdS-Cauchy surface $\Sigma$ on which $m(A)$ is the minimal surface. Let $N$ be the surface formed by following the $\ell$-orthogonal null congruence from $\sigma$ toward the future and by also following the $k$-orthogonal null congruence from $\sigma$ toward the past, with the convention that a given geodesic remains in $N$ only so long as it lies on the boundary of the future/past of $\sigma$. Define $\tilde{\sigma}$ as the intersection of $N$ with $\Sigma$, $N\cap \Sigma$. Now, the future directed portion of $N$ has $\theta \le 0$ at $\sigma$, while the past directed portion has $\theta=0$ at $\sigma$. The null curvature condition implies that null rays can only focus as they move away from $\sigma$, decreasing the the total area of $N$. As in the discussion of theorems 3.1.1 and 3.1.2, either condition (i) or (ii) suffices to guarantee that the flux of area through any regulator surface $z=z_0$ vanishes as $z_0 \rightarrow 0$, and otherwise we discuss only the coefficient of the leading divergence. Since $m(A)$ is minimal on $\Sigma$ we thus find $\text{Area}_{\text {ren}} (\sigma) \ge \text{Area}_{\text {ren}}(\tilde{\sigma}) \ge \text{Area}_{\text {ren}}(m(A))$.


\section{Divergences}
\label{sec:div}
Entanglement entropy and Causal Holographic Information are both infinite, as are the areas of the boundary anchored surfaces that measure them. In particular, it has been shown \cite{FreivogelMosk:2013} that subleading divergences in the area of the causal surface generally differ from those of extremal surface: while the entanglement divergences can be written as the integral of local geometric quantities on $\partial A$, subleading divergences of the causal surface generally cannot. However, \cite{FreivogelMosk:2013} conjectured that the leading-order divergences agree for $d>2$. We investigate the analogous issues below for marginally trapped surfaces $\sigma$ anchored to $\partial A$, showing first that leading area-divergence of $\sigma$ does in fact agree with that of $m(A)$, and then demonstrating that subleading divergences generally differ.

\subsection{Leading Order Divergences}
\label{sec:leading}
It is useful to begin with the Fefferman-Graham expansion of the metric \cite{FeffermanGraham:1985}. In $d \ge 2$ dimensions, this takes the form
\begin{equation}
ds^2 = g_{ab} dx^a dx^b = \frac{l^2}{z^2}(dz^2 + \tilde{\gamma}_{ij}(x,z)dx^idx^j), 
\end{equation}
where $l$ is the AdS length scale, $x$ ranges over the boundary coordinates, and\footnote{At least when there are no operators or non-metric sources with conformal dimension $d \geq \Delta \geq 0$}
\begin{equation}
\tilde{\gamma}_{ij}(x,z)= \tilde{\gamma}_{ij}^{(0)}(x) + z^2 \tilde{\gamma}_{ij}^{(2)}(x) + ...z^d\left(\tilde{\gamma}_{ij}^{(d)}(x) + \bar{\tilde{\gamma}}_{ij}^{(d)}(x)\text{log}(z^2)\right).
\end{equation}
Here, $\tilde{\gamma}_{ij}^{(0)}$ is the metric on the boundary, and the logarithmic term only appears for even $d$. Note that $ds^2 =  \frac{l^2}{z^2}(dz^2 + \tilde{\gamma}_{ij}^{(0)}(x)dx^idx^j) + O(z^0)$. In particular, the unphysical conformally-rescaled metric 
\begin{equation}
\label{eq:rescaledmetric}
\widetilde{ds}^2 =  \tilde g_{ab} dx^a dx^b =  \frac{z^2}{l^2} ds^2 
\end{equation}
is finite as $z \rightarrow 0$ and gives the bulk the structure of a manifold $\tilde M$ with boundary.

Consider any marginally-trapped surface $\sigma$ whose derivatives that are $C^1$ in $\tilde M$. Then tangent vectors to $\sigma$ are well-defined both on $\tilde M$ and on the AlAdS boundary, and the geodesic equation on $\sigma$ is also well-defined. We expect this condition to hold for surfaces constructed as in Section \ref{sec:prelim}. 

Near the boundary, we can use (generalized) Riemann normal coordinates $\{\hat x^\alpha \} = \{\hat x^i, z\}$ on $\sigma$ defined by using the unphysical metric \eqref{eq:rescaledmetric} to construct a congruence of geodesics orthogonal to the AlAdS boundary. Here we have generalized the notion of Riemann normal coordinates slightly by not requiring $z$ to be proper distance.   The $\hat x^i$ are constant along the geodesics and agree with $x^i$ on the AlAdS boundary; however, they do not generally agree with the Fefferman-Graham $x^i$ in the interior.   

In terms of the coordinates $\hat x^\alpha$ on $\sigma$, the tangents to the above geodesics are $\tau^\alpha \partial_\alpha = \partial_z|_{\hat x^i}$.  As a result, the metric induced by \eqref{eq:rescaledmetric} takes the form
\begin{equation}
\label{eq:inducedsigmametric}
\tilde h_{\alpha \beta} dx^\alpha dx^\beta = \tilde q_{ij}(z) d\hat x^i d\hat x^j +  \widetilde{|\tau|}^2 dz^2,
\end{equation}
with $\tilde q_{ij}(z) = \tilde q^{(0)}_{ij} + O(z)$ for $\tilde q^{(0)}{}_i^j$ the projector onto the anchor set $\partial A$ and 
where the error term allows for $\partial A$ to have non-vanishing extrinsic curvature in $\sigma$ (as computed with respect to \eqref{eq:rescaledmetric}) even though the extrinsic curvature of the full AlAdS boundary vanishes with respect to this metric. We remind the reader that we use $\tilde{\gamma}^{(0)}_{ij}$ to raise and lower $i,j$ indices on the boundary. In \eqref{eq:inducedsigmametric}, $\widetilde{|\tau|}^2$ is the norm of $\tau^\alpha$ in the rescaled metric \eqref{eq:rescaledmetric}.  Note that while $\sigma$ is spacelike in the bulk, a priori the norm $\widetilde{|\tau|^2}$ might vanish as $z \rightarrow 0$.  

Since $\sigma$ is marginally trapped, it has $\theta_k =0$.  We are free to choose $k$ to have Fefferman-Graham components $k^a$ that vanish like $z$ as $z \rightarrow 0$ so that $\tilde k^a : = k^a/z$ remains finite. We also define a rescaled extrinsic curvature tensor
$\tilde K_{abc}$  such that for any null vector field $v^a$ orthogonal to $\sigma$ we have  
\begin{equation}
v^a \tilde K_{a bc} : =  z v^aK_{a bc} = \frac{1}{2} z \pounds_v \left(\frac{l^2}{z^2} \tilde h_{bc} \right),
\end{equation}
where $\pounds_v$ denotes the Lie derivative along $v$ and $h_b^c$ is the projector onto $\sigma$.
Then the condition $\theta_k =0$ is then equivalent to $\tilde k^a \tilde K_{a bc} \frac{l^2}{z^2}\tilde h^{bc} = 0$, where $\tilde h^{bc}$ is obtained from $\tilde h_{bc}$ by raising indices with $\tilde g^{ab}$. Note that converting \eqref{eq:inducedsigmametric} to Fefferman-Graham coordinates $\{x^a\} = \{x^i,z\}$ gives 
\begin{equation}
\label{eq:inducedsigmametric2}
\tilde h_{ab}  = \tilde q_{ab} +  \frac{\tilde \tau_a \tilde \tau_b dx^a dx^b}{\widetilde{|\tau|}^2},
\end{equation}
where $\tilde \tau_a = \tilde g_{ab} \tau^b$ and $\tilde q_{ab} = \tilde g_{ac} \tilde q_b{}^c$, where $\tilde q_b{}^c$ is the projection onto the space tangent to $\sigma$.
We may therefore compute as follows:
\begin{eqnarray}
\label{eq:ComputeK}
2 \tilde k^a \tilde K_{a bc}  &=& z \pounds_{\tilde k} \left(\frac{l^2}{z^2}  \tilde h_{bc} \right)  \\ &=& \frac{l^2}{z}  \pounds_{\tilde k} \tilde q_{bc} - 2l^2\frac{\pounds_{\tilde k} z}{z^2} \tilde q_{bc}  -  l^2z^{-1}\left( \pounds_{\tilde k}  \ln \widetilde{|\tau|}^2\right) \frac{\tilde \tau_a \tilde \tau_b }{\widetilde{|\tau|}^2} + \frac{l^2}{z\widetilde{|\tau|}^2} \left(  \tilde \tau_a \pounds_{\tilde k} \tilde \tau_b  + \tilde \tau_b \pounds_{\tilde k} \tilde \tau_a  \right). \ \
\end{eqnarray}

Since $\tau^a$ is annihilated by $\tilde q_{ab}$, we can derive the useful relation
\begin{equation}
\tau^b \tau^c \pounds_{\tilde k} \tilde q_{bc}  = \tau^b  \pounds_{\tilde k} ( \tau^c \tilde q_{bc}) -\tau^b \tilde q_{bc} \pounds_{\tilde k}  \tau^c  =0.
\end{equation}
Contracting \eqref{eq:ComputeK} with $\frac{l^2}{z^2}\tilde h^{bc}$  thus yields
\begin{eqnarray}
\label{eq:contract1}
0 = 2\frac{l^2}{z^2}\tilde h^{bc} \tilde k^a \tilde K_{a bc}  &=& z \pounds_{\tilde k} \left(\frac{l^2}{z^2}  \tilde h_{bc} \right)  \\ &=& O(z)  -2 (d-1) \left( \pounds_{\tilde k} z \right)   - z \left( \pounds_{\tilde k}  \ln \widetilde {|\tau|}^2\right)  + \frac{2 z}{\widetilde{|\tau|}^2} \tau^b \pounds_{\tilde k} \tilde \tau_b.
\end{eqnarray}
But we also find
\begin{eqnarray}
\tau^a \pounds_{\tilde k} \tilde \tau_a &=& \tau^a \pounds_{\tilde k} (\tilde g_{ab} \tau^a) = \frac{1}{2} \pounds_{\tilde k} \widetilde{|\tau|}^2 - \frac{1}{2}\tau^a \tau^b \pounds_{\tilde k} \tilde g_{ab} \\
&=& 
\frac{1}{2} \pounds_{\tilde k} \widetilde{|\tau|}^2 - \tau^a \tau^b \tilde \nabla_a \tilde k_b 
= \frac{1}{2} \pounds_{\tilde k} \widetilde{|\tau|}^2 - \tau^a \tilde \nabla_a (\tau^b \tilde k_b) = \frac{1}{2} \pounds_{\tilde k} \widetilde{|\tau|}^2,
\end{eqnarray}
where $\tilde \nabla_a$ is the covariant derivative for $\tilde g_{ab}$ and the steps on the last line follow from the (non-affinely parametrized) geodesic equation $\tau^b \tilde \nabla_b \tau^a \propto \tau^a $ and the orthogonality of $\tau_a$ and $\tilde k^a$. As a result, \eqref{eq:contract1} yields
\begin{equation}
\label{eq:contract2}
0 = \frac{l^2}{z^2}\tilde h^{bc} \tilde k^a \tilde K_{a bc} =   -2 (d-1) \left( \pounds_{\tilde k} z \right) + O(z)
= -2(d-1) \tilde k^z + O(z).   
\end{equation}
I.e., to this order $\tilde k$ is tangent to the AlAdS boundary and is thus a null normal to $\partial A$ with respect to the boundary metric $\tilde{\gamma}^{(0)}_{ij}.$ 

On the other hand, since $\tilde k$ is orthogonal to $\sigma$ we have $\frac{\tilde k_a \tau^a}{|\tau|^2} = 0$, so $\tau_a$ lies in the null plane defined by 
$\tilde k_a$.  And since $\tau^a$ is by definition orthogonal to $\partial A$, we may use the $C^1$ nature of $\tau$, the fact that
$\tau^\alpha \partial_\alpha = \partial_z|_{\hat x^i}$, and the equality of the
$z$-component of $\tau$ in $\{\hat x^\alpha \}$ coordinates with that in Fefferman-Graham coordinates
to write $\tau^{a} = \tau^k \tilde k_a + \partial_z x^a + O(z)$ for some finite coefficient $\tau^k$ and where the $O(z)$ term is again orthogonal to $\tilde k$. Since $\tilde k^a$ is both null and (to order $z$) orthogonal to $\partial_z$, we find $\widetilde{|\tau^2|} = 1 + O(z)$; in particular, $\tau^a$ remains spacelike at $z=0$.  

We are now ready to calculate the leading area-divergence of $\sigma$. This is simplest in the coordinates $x^\alpha = \{\hat {x}^i, z\}$ where $\tau^\alpha \partial_\alpha = \tau^z \partial_z|_{\hat {x}^i}$ and the metric induced by $\tilde g_{ab}$ is \eqref{eq:inducedsigmametric}. It is clear that the physical area of $\sigma$ takes the form
\begin{equation}
\text{Area}[\sigma] = \int_{\partial A} d^{d-2} x \int dz \frac{l^{d-1}\sqrt{q^{(0)}}}{z^{d-1}} + O(z^{-(d-1)}),
\end{equation}
where the leading term agrees with the leading area-divergence for an extremal surface anchored to $\partial A$ (as it must, since an extremal surface is also marginally-trapped and we have shown this term to be the same for all marginally-trapped surfaces).

Though we leave the details for future work, since we found $\widetilde{|\tau^2|} = 1 + O(z)$ but used only $\lim_{z\rightarrow 0} \widetilde{|\tau^2|} = 1$ it seems clear that this result also extends to marginally-trapped surfaces which are more singular at the boundary.  This plausibly includes all cases where the constraints $F,G$ of Section \ref{sec:prelim} admit expansions in fractional powers of $z$.


\subsection{Subleading Divergences}
While the are of $\sigma$ agrees with that of an extremal surface to leading order, the subleading divergences do not generally match. We can show this by example. Consider the $d+1=5$ dimensional bulk metric,
\begin{equation}
ds^2 = \frac{1}{z^2}(-dt^2+dz^2 +dy^2+ X(x,y,u)dx^2 + du^2),
\end{equation}
where $X(x,y,u)$ is an arbitrary function.  We will take our boundary region to be a strip with $y\in \left[-f_0,f_0\right]$ and $t=g_0$ and take the constraints to be $t - G(z)=0$ and $y - F(z)=0$. We can take both the constraints and the metric to be expandable in power series: 
\begin{eqnarray}
F(z)&=&f_0+ f_1 z+ f_2 z^2+f_3 z^3+...,\\
G(z)&=&g_0+ g_1 z+g_2 z^2 +g_3 z^3+...,\\
X(x,y,u)&=& X_0(x,u)+ F(z) X_1(x,u)+ F^2(z) X_2(x,u)+ F^3(z) X_3(x,u)+...,
\end{eqnarray}
for some functions $X_i$ and constants $f_i, g_i$. A calculation shows that for any $F$ and $ G$ the two $\sigma$-orthogonal null congruences have
\begin{eqnarray}
\theta_{\pm }=\pm \frac{3 z \left(f_1 g_1^2 -f_1^3-f_1 \pm g_1 \sqrt{f_1^2-g_1^2+1}\right)}{\left(f_1^2+1\right) \sqrt{f_1^2-g_1^2+1}} + O(z^2).\\
\end{eqnarray}
Choosing $\theta_{+}=0$ would then impose $f_1= g_1$, while $\theta_- = 0$ would impose $f_1=-g_1$. We can similarly solve $\theta_+=0$ or $\theta_-=0$ to second order, and we find 
\begin{equation}
f_2 =  \frac{f_0^2 X_2(y,u)+2 f_0 X_1(y,u)+3 X_0(y,u)}{8 f_0\left(f_0^3 X_3(y,u)+f_0^2 X_2(y,u)+ f_0X_1(y,u)+X_0(y,u)\right)}\pm g_2-\frac{3}{8 f_0}.
\end{equation}

Now, the area of the marginally trapped surface will be given by
\begin{equation}
A = \int \text{d}z\text{d}u \text{d}x \frac{1}{z^3}\sqrt[]{1 +F'(z)^2 - G'(z)^2} \, \sqrt[]{X(x,y,u)}.
\label{eq:this}
\end{equation}
Evaluating (\ref{eq:this}) on our solutions for $f_1$ and $f_2$, yields
\begin{eqnarray}
\begin{aligned}
A =& \int \text{d}z\text{d}u\text{d}x \frac{1}{z^3}\left(\sqrt{X_0(x,u)+f_0 X_1(x,u)+f_0^2 X_2(x,u)+f_0^3  X_3(x,u)}\right)\\
&+ \int \text{d}z\text{d}u\text{d}x\frac{1}{z^2}\left(\frac{g_1 X_1(x,u)+2 f_0 g_1 X_2(x,u)+ 3 f_0^2 g_1 X_3(x,u)}{4 \sqrt{X_0(x,u)+f_0 X_1(x,u)+f_0^2 X_2(x,u)+f_0^3 X_3(x,u)}}\right)+ O(\text{ln}z).
\end{aligned}
\end{eqnarray}
As expected, the leading divergence is fixed by boundary conditions only, as seen by the fact that it depends only on $f_0$. The subleading divergence, however, depends on $g_1$ as well. This $g_1$ is the asymptotic slope of the slice determined by $G(z)$, and thus will generally differ from that of the the extremal surface.

Since the divergence depends on the slope of the slice, we expect it to have some relation to the tangent plane to the marginally trapped surface. Consider the unique tangent vector
\begin{equation}
\tau^a = (t,x,y,u,z) = \frac{\partial}{\partial z} (G(z),x,F(z), u, z).
\end{equation}
that is orthogonal to the boundary of $\sigma$. We expect the divergence to be in part determined by $\tau^a$, though it must be contracted with some one index object that contains information about the boundary region $A$. A natural candidate is the trace of the extrinsic curvature of $\partial A$,
\begin{equation}
{K^{(b), i}}_{i}  = \nabla^i {n^{(A)}}_i,
\end{equation}
where the $i$ index runs over the boundary indices,  the $b$ index runs over the 2 vectors orthogonal to our boundary subregion (and contained in the boundary), and $n$ is the normal to $\partial A$ that points outwardly away from $A$. The only nonzero component is 
\begin{equation}
{K^{(y), i}}_{i} = \frac{X_1(x,u)+2 X_2(x,u)y+3 X_3(x,u)y^2+...}{2 \left(X_0(x,u)+X_1(X,u)y+X_2(x,u)y^2 +3 X_3(x,u)y^3+...\right)}. 
\end{equation}

Define ${\tau}_\parallel$ to be the projection of $\tau^a$ into the AdS boundary. We can then contract with $K =(K^t,0,K^y,0) = (0, K^y,0)$. This gives
\begin{equation}
K\cdot {\tau}_\parallel = \frac{\left(X_1(x,u)+2 X_2(x,u)y+3 y^2 X_3(x,u)+...\right)f_1}{2\left(X_0(x,u)+ X_1(x,u)x +X_2(x,u)y^2 + X_3(x,u)y^3+...\right)}.
\label{eq:extrinsicdottau}
\end{equation}
Integrating (\ref{eq:extrinsicdottau}) over $\partial A$ and using $y=f_0$  gives
\begin{eqnarray}
\begin{aligned}
&\int_{\partial A} \text{d}u\text{d}x \ \sqrt[]{X(x,y,u)} K \cdot {\tau}_\parallel\\
 = &\int \text{d}u\text{d}x \frac{g_1 \left(X_1(x,u)+2 f_0 X_2(x,u)+3 f_0^2 X_3(x,u)+...\right)}{2 \sqrt{X_0(x,u)+f_0 X_1(x,u)+f_0^2 X_2(x,u) +f_0^3 X_3(x,u)+...}},\\
\end{aligned}
\end{eqnarray}
so that 
\begin{equation}
\begin{aligned}
A &= \int \text{d}z \frac{1}{z^3} \int_{\partial A} \text{d}u\text{d}x \ \sqrt[]{X(x,y,u)} K \cdot {\tau}_\parallel\\
&= \int_{A} \sqrt[]{h_A} \left( -\frac{1}{2z^2} - \frac{1}{z} K \cdot   {\tau}_\parallel\right).
\end{aligned}
\end{equation}
for $h_A$ the induced metric on $A$ from the boundary metric. Thus, the subleading divergence is given by the integral of the trace of the  extrinsic curvature of $\partial A$ contracted with the tangent vector orthogonal to the boundary of the marginally trapped surface. 

\section{Thermodynamics}
\label{sec:thermo}
It has been previously shown that, when they are compact, the areas of leaves of holographic screens monotonically increase \cite{EngelhardtBousso:2015,EngelhardtBousso:2015-2}. In this section, we generalize this proof to the case of non-compact leaves. The main difficulty in the original proof is constraining the ways in which holographic screens can change from spacelike to timelike. If, for instance, we knew that flowing along our screen moved a given leaf only toward the past and toward the boundary, we could quickly conclude that the area increased. If it was toward the past, we could first flow infinitesimally to the past along the $k$-congruence (i.e., in the negative $k$ direction), and then to the past along the $\ell$-congruence (i.e., in the negative $k$ direction). Along the $k$-congruence, the area remains constant to first order as one moves away from any leaf. Since the expansion is non-negative in the negative $\ell$ direction, to first order the area cannot decrease. The net change is then non-negative, and the area of the leaves will not decrease. Likewise, if the nearby leaf was spacelike and towards the boundary, we could first flow along the future $\theta_k=0$ direction, then along the past $\theta_\ell>0$ direction, leading again to non-decreasing area. Reversing these arguments, if the nearby leaf were to the future or spacelike away from the boundary, the area would decrease.

Before we rule out out the problematic flow directions, we will review the assumptions of \cite{EngelhardtBousso:2015-2}, and those made here. 

\textit{Definition:} We can define a set of leaf-orthogonal curves {$\gamma$} such that every point $p$ in our holographic screen $H $ lies on one curve. We can further choose a parameter $r$ that is constant along each leaf $\sigma$ but increases monotonically along each curve $\gamma$. 

Since $\gamma$ is taken to be orthogonal to each leaf, its tangent vector $h^\mu$ can be written as a linear combination of the null congruences, 
\begin{equation}
h^\mu  = \alpha \ell^\mu + \beta k^\mu.
\end{equation}
where $h$ is normalized such that $r$ increases at unit length along $h$. Note that $\alpha$ and $\beta$ cannot be both zero, though they may approach zero at the AlAdS boundary.  

We will then use the following assumptions about the spacetime, following \cite{EngelhardtBousso:2015-2}. As above, we assume the null curvature condition, $R_{ab}= k^ak^b\geq0$. We also assume two generic conditions. One, that $R_{ab}k^ak^b + \xi_{ab}\xi^{ab}>0$ at every point on our holographic screen for the $k$-directed congruence. Two, if we denote by $H_0, H_+,$ and $H_-$ the sets where respectively $\alpha=0, \alpha>0$, and $\alpha<0$ on $H$, then $H_0 = \partial H_- = \partial H_+$. Further, we assume that every inextendible portion of our holographic screen is either entirely timelike, or contains a complete leaf. Finally, we assume that every leaf $\sigma$ on our screen splits a Cauchy surface $\Sigma$ into two disjoint components. The extent to which such assumptions are reasonable for compact leaves is discussed in \cite{EngelhardtBousso:2015-2}; similar comments apply here. From these assumptions, it follows that at least one leaf will have definite sign of $\alpha$. We can take $r=0$ on this leaf, and orient $r$ such that $\alpha<0$. It suffices to consider each connected component of $H$ separately, so we may take $H$ to be connected for the rest of the argument. 

We make one additional assumption beyond those of \cite{EngelhardtBousso:2015-2}, namely that $H$ can be deformed continuously into a sequence of screens $H_a$ by deforming the anchor sets of each leaf $\sigma_a(r_i)$ in a spacelike direction such that $ A(\sigma_a(r_i)) \supset A(\sigma_a(r_j))$ for $r_i<r_j$, and $A(\sigma_a(r_i)) \subset A(\sigma_b(r_i))$ for $a<b$, with $H_0=H$. 

We can now quickly reduce our setting to (almost) the one considered in \cite{EngelhardtBousso:2015-2}. We proceed by first recalling that, as discussed above, the essence of the argument is really a theorem about certain changes of sign as one moves along the holographic screen. Those signs are conformally invariant, as they do not depend on the metric. So it suffices to prove the `restricted changes of sign' version of the theorem for our screen as embedded in the unphysical conformally-rescaled spacetime associated with the metric $\widetilde{ds}^2$ of \eqref{eq:rescaledmetric}. The area increase theorem then follows for the original screen in the physical spacetime by using the conditions $\theta_k=0, \theta_\ell \le 0$ that hold there.

In this unphysical spacetime the leaves are now compact, but they have boundaries at the AlAdS boundary. To reduce this to the no-boundary case considered in \cite{EngelhardtBousso:2015-2}, we now consider two copies of the unphysical conformally rescaled spacetime and identify them along their AlAdS boundaries. The resulting ${\mathbb Z}_2$-symmetric spacetime is compact, globally-hyperbolic (in the usual non-AdS sense), and has no boundary. This procedure also glues together the two copies of $H_a$ and $H$ to make holographic screens with compact leaves. 

The only remaining difference from the setting of \cite{EngelhardtBousso:2015-2} is that, in the doubled spacetime, the leaves are generally only continuous and may not be smooth. However, the proof of \cite{EngelhardtBousso:2015-2} proceeds by firing null congruences from various leaves and studying their intersections (or, at least the intersection of the associated boundaries of future and/or past sets) with the screen. Having shown that these intersections lie entirely on one side of the $r=0$ leaf, continuity and compactness guarantee the intersection to have a minimum (or maximum) $r$. Smoothness is then used to argue that the intersection it tangent to the leaf at this minimum (maximum) $r$, and to find a contradiction with our Corrollary 2.3.  In our case, taking $H_a$ to be small deformations of $H$ satisfying the above conditions guarantees that there can be no intersection on the AlAdS boundary, and so in particular the minimum (maximum) $r$ does not occur there. Since the doubled screen is smooth away from the AlAdS boundary, the rest of the argument then proceeds as in \cite{EngelhardtBousso:2015-2} to yield:

\textit{Theorem 5.1} Let $H$ be a future holographic screen satisfying the above assumptions, with a leaf orthogonal tangent vector field $h^a  = \alpha \ell^a + \beta k^a$. Then $\alpha \leq 0$ on all of $H$.

The desired result then follows immediately.

\textit{Theorem 5.2} The area of the leaves of $H$ increases monotonically as measured by the physical bulk metric $ds^2$.

\section{Discussion}
\label{sec:disc}
We have shown that boundary-anchored holographic screens anchored in AlAdS spacetimes have several interesting properties. First, for a boundary ``spatial region'' (partial Cauchy surface) $A$ with no extremal surface barriers between the screen and the compliment of $A$ in the boundary Cauchy surface, any screen anchored to $\partial A$ lies above the future horizon of $D_{bndy}(A)$ but inside of the entanglement wedge of $A$. We further showed that the area of the holographic screen is bound below by the area of the extremal surface, and in certain cases, bounded above by the quantity we called future causal holographic information (fCHI) defined by the area of a cut of the causal horizon.

We also studied the divergences in area of the holographic screens. While the leading divergence of a holographic screen matches that of the extremal surface, the first subleading divergence generally differs from that of extremal surfaces. Finally, we have shown that, under a continuous choice of flow along leaves, there is a monotonic change in area, generalizing the results of \cite{EngelhardtBousso:2015-2} to the case of non-compact leaves. 

A technical complication in our work is the large set of assumptions (matching those of \cite{EngelhardtBousso:2015-2}) used to prove the 2nd law in Section \ref{sec:thermo}. Most of these are clearly true in the generic case, but this is far from clear for the assumption that every inextendible portion of the screen is either entirely timelike or contains a complete leaf. Another complication is that the various bounds on the area of boundary-anchored marginally-trapped surfaces are generally useful only for surfaces already known to coincide with extremal surfaces up to corrections vanishing faster than $z^{d-2}$ -- the order required to make the area only finitely different from that of an HRT surface. It would be much more natural to find a simple construction of marginally-trapped surfaces for which these assumptions were guaranteed to be satisfied, or which forced the desired results to apply more generally. It would also be interesting to (perhaps numerically) explore whether the inextendible portion assumption holds in general, either in our boundary-anchored setting or in the original compact context of \cite{EngelhardtBousso:2015-2}. However, we leave such explorations for future work.

Since the holographic screen lies inside the entanglement wedge, it should describe some property of any dual field theory in $D_{bndy}(A)$. Interestingly, this differs substantially from the original conjecture of \cite{Bousso:1999cb} regarding the holographic properties of such screens which took the screen to describe degrees of freedom on what we would call the `inside' (i.e., the $\ell$-congruence side) of the screen. In contrast, our result suggests the screen to describe properties of $D_{bndy}(A)$ and the associated part of the entanglement wedge `outside' the screen (i.e., on the $k$-congruence side). As described in Section \ref{sec:order}, the above-mentioned bounds on the area of any leaf suggest such areas to measure a coarse-grained entropy for the dual CFT (though one that is finer-grained than that associated with fCHI).

Indeed, while this work was in preparation, ref. \cite{EngelhardtWall:2017} appeared which studied a closely related issue. Their work shows that  the area of a black hole’s apparent horizon measures a coarse-grained entropy, where the coarse-graining is over all solutions in the interior, keeping the geometry of the exterior fixed. In particular, they show that the apparent horizon area agrees with that of the largest HRT surface consistent with the above constraints. Although \cite{EngelhardtWall:2017} does not study boundary anchored surfaces, we anticipate it to admit an extension to boundary anchored leaves whose divergences match those of extremal surfaces. In contrast, however, the analogous result is clearly forbidden when the divergences of the leaf fail to match all state-independent divergences of the extremal surface. 

Now, as in \cite{EngelhardtBousso:2015}, the area-increase result of Section \ref{sec:thermo} suggests a thermodynamic interpretation for the area. Here we find that the area increases toward the boundary, in the sense that one moves in the direction along the holographic screen that is most closely associated with the $k$-congruence, when the screen moves in a spacelike direction. Interestingly, on a timelike part of the screen this corresponds to moving the leaf toward the past \cite{EngelhardtBousso:2015}. There is also the somewhat uncomfortable property that the area-increase theorem requires comparing entire leaves; deforming a cut of the screen locally toward the future (so that it no longer coincides with a leaf) is not generally guaranteed to increase the area.

Recall, however, that Section \ref{sec:order} noted that the area of a leaf is also bounded above by the area of a cut $Y$ of the future horizon when the leaf is constructed by requiring it to lie in the boundary of the future $\dot{I}^+(S)$ of some set $S$ in the AlAdS boundary satisfying $\partial S = \partial A$. This $Y$ is the intersection of the future horizon with $\dot{I}^+(S)$ introduced in \cite{Kelly:2013aja}, and its $A/4G$ is naturally called future causal holographic information. Here we again emphasize the difference in perspective from constructions of marginally-trapped surfaces from light cones in \cite{Bousso:1999cb}, as the light cones of \cite{Bousso:1999cb} were generated at $\sigma$ by the $k$-congruence while our $\dot{I}^+(S)$ is generated at $\sigma$ by the $l$-congruence.  

Since deformations of $S$ toward the future now move $\dot{I}^+(S)$ outward when the screen is spacelike, in such cases our second law makes the associated areas of marginally-trapped surfaces monotonically non-decreasing under any such flow. This reinterpretation of the results of Section \ref{sec:thermo} would then remove the discomforts mentioned above. In particular, when the screen is spacelike we now find non-increase toward what is clearly the future and, in addition, the system may be pushed forward in time independently at each point. We therefore hope to investigate this construction further in the future, as always with an eye toward better understanding the interpretation in the dual CFT.

Finally, as always in such discussions, one would like to progress beyond leading order in the bulk semi-classical expansion. This would presumably involve replacing the area of each leaf with the generalized entropy as in \cite{Faulkner:2013ana,EngelhardtWall:2013,Jafferis:2015del,Dong:2017xht}. However, it is unclear just how the bulk entanglement term should be defined for bulk gravitons. While the arguments of \cite{Jafferis:2015del} and \cite{Dong:2017xht} can be used to define this entanglement across an HRT surface, at least at present there is no general understanding of how to define such entanglement across a general bulk surface -- or even a general one that is marginally trapped. The issue is a classic one associated with the failure of the linearized graviton action to be gauge invariant on off-shell backgrounds such as those that would naturally be used in attempting to define this entanglement using the replica trick. Nevertheless, it would still be natural to explore the effects of entanglement terms associated with other bulk fields while awaiting a better understanding of graviton entanglement.

\section*{Acknowledgements}
It is a pleasure to thank Raphael Bousso, Netta Engelhardt, and Aron Wall for useful conversations. This work was supported in part by a U.S. National Science Foundation under grant number PHY15-04541 and also by the University of California. BG-W was also supported in part by a National Science Foundation Graduate Research Fellowship.

\end{document}